\documentclass[twocolumn]{aastex701}
\submitjournal{PASP}

\newcommand{\topic}[1]{\vspace{0.5ex}\noindent\textbf{#1}}

\begin{document}

\title{Winding, Unwinding, Rewinding the Gaia Phase Spiral}

\author[]{Neige Frankel}
\affiliation{Canadian Institute for Theoretical Astrophysics, University of Toronto, 60 St. George Street, Toronto, ON M5S 3H8, Canada}
\email{frankel@cita.utoronto.ca}

\author[]{Marcin Semczuk}
\affiliation{Institut de Ciències del Cosmos (ICCUB), Universitat de Barcelona (UB), c. Martí i Franquès, 1, 08028 Barcelona, Spain}
\affiliation{Departament de Física Quàntica i Astrofísica (FQA), Universitat de Barcelona (UB), c. Martí i Franquès, 1, 08028 Barcelona, Spain}
\affiliation{Institut d'Estudis Espacials de Catalunya (IEEC), c. Gran Capità, 2-4, 08034 Barcelona, Spain}
\email{msemczuk@fqa.ub.edu}

\author[]{Teresa Antoja}
\affiliation{Institut de Ciències del Cosmos (ICCUB), Universitat de Barcelona (UB), c. Martí i Franquès, 1, 08028 Barcelona, Spain}
\affiliation{Departament de Física Quàntica i Astrofísica (FQA), Universitat de Barcelona (UB), c. Martí i Franquès, 1, 08028 Barcelona, Spain}
\affiliation{Institut d'Estudis Espacials de Catalunya (IEEC), c. Gran Capità, 2-4, 08034 Barcelona, Spain}
\email{tantoja@fqa.ub.edu}

\author[]{Sukanya Chakrabarti}
\affiliation{Department of Physics \& Astronomy, University of Alabama, Huntsville, 35899}
\email{sc0236@uah.edu}

\author[]{Rimpei Chiba}
\affiliation{Canadian Institute for Theoretical Astrophysics, University of Toronto, 60 St. George Street, Toronto, ON M5S 3H8, Canada}
\affiliation{Department of Astronomy, The University of Tokyo, 7-3-1 Hongo, Bunkyo-ku, Tokyo, 113-0033, Japan}
\email{rimpei-chiba@g.ecc.u-tokyo.ac.jp}

\author[]{Robert J. J. Grand}
\affiliation{Astrophysics Research Institute, Liverpool John Moores University, 146 Brownlow Hill, Liverpool, L3 5RF, UK}
\email{R.J.Grand@ljmu.ac.uk}

\author[]{Jason A. S. Hunt}
\affiliation{School of Mathematics \& Physics, University of Surrey, Stag Hill, Guildford, GU2 7XH, UK}
\email{j.a.hunt@surrey.ac.uk}

\author[]{Sergey Khoperskov}
\affiliation{Leibniz-Institut für Astrophysik Potsdam (AIP), An der Sternwarte 16, 14482 Potsdam, Germany}
\email{sergey.khoperskov@gmail.com}

\author[]{Zhao-Yu Li}
\affiliation{Department of Astronomy, School of Physics and Astronomy, Shanghai Jiao Tong University, 800 Dongchuan Road, Shanghai 200240, China}
\affiliation{Key Laboratory for Particle Astrophysics and Cosmology (MOE) / Shanghai Key Laboratory for Particle Physics and Cosmology, Shanghai 200240, China}
\email{lizy.astro@sjtu.edu.cn}

\author[]{Artem Lutsenko}
\affiliation{Dipartimento di Fisica e Astronomia, Universitá di Padova, Vicolo Osservatorio 3, I-35122 Padova, Italy}
\affiliation{INAF - Padova Observatory, Vicolo dell'Osservatorio 5, I-35122 Padova, Italy}
\email{artem.lutsenko@studenti.unipd.it}

\author[]{Pau Ramos}
\affiliation{National Astronomical Observatory of Japan, Mitaka, Tokyo 181-8588, Japan}
\email{pau.ramos@nao.ac.jp}

\author[]{Kiyan Tavangar}
\affiliation{Department of Astronomy, Columbia University, New York, NY 10027, USA}
\email{k.tavangar@columbia.edu}

\author[]{Lawrence M. Widrow}
\affiliation{Department of Physics, Engineering Physics, and Astronomy, Queen's University, Kingston, Canada, K7L 3N6}
\email{widrow@queensu.ca}

\begin{abstract}

The Gaia Space Satellite has transformed the field of Galactic Dynamics by collecting 6D phase space information for hundreds of millions of stars. In 2018, it enabled the discovery of the Gaia Phase Spiral \citep{antoja_2018}, a clear signal in the vertical 
motion of the stars that reveals how far from equilibrium the Galactic disk is. Seven years after the discovery of this structure, a workshop dedicated to the Phase Spiral took place at the Lorentz Center. Workshop participants summarized the current state of knowledge about the Phase Spiral and identified open questions and key areas to continue progressing in understanding the origin of the Phase Spiral and the physics governing the response of the disk to perturbations. Here, we aim to summarize the content and discussions of this workshop, share the resources that have been produced at this workshop with the broader community, and invite interested individuals to join on the projects that started.

\end{abstract}

\keywords{\uat{Milky Way Galaxy}{1051}, \uat{Mily Way Dynamics}{1054}, \uat{Milky Way Evolution}{1052}}



\section{Introduction\label{sec:intro}}
\begin{figure*}
    \centering
    \includegraphics[width=\textwidth]{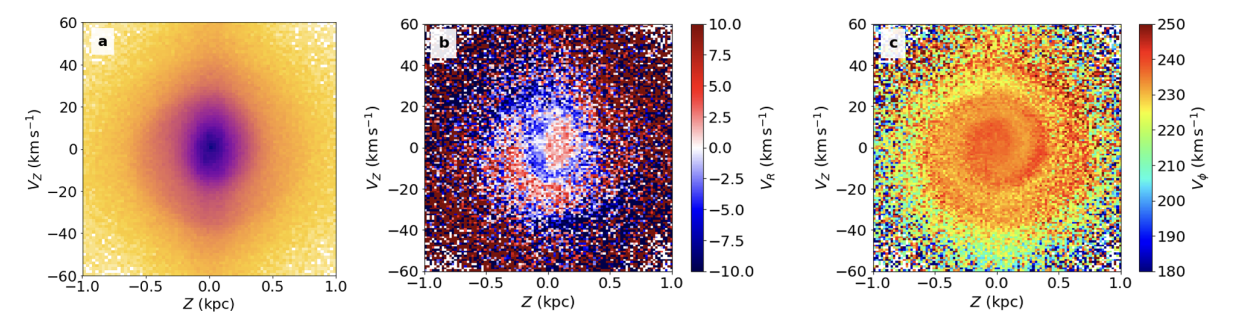}
    \caption{Phase spiral from \cite{antoja_2018}, reproduced with permission. Three panels present the distributions of stars in the Solar neighborhood in the $Z-V_Z$ plane, where $Z$ is the height above and below the plane of the Galactic disk and $V_Z$ is the velocity in that direction. Panel a) shows the 2D histogram of stars. Panel b) is colored by the median cylindrical radial velocity $V_R$ and panel c) by the median azimuthal velocity $V_\varphi$.}
    \label{fig:antoja}
\end{figure*}

Galactic dynamics shapes galaxies during their secular evolution and is central to understanding galaxy formation. Historically, theoretical work focused on near-equilibrium states and observations were limited to integrated light from external galaxies, or sparse kinematic data in the Solar-Neighborhood.
{\cite{oort_1932} first estimated the force normal to the midplane and the local density from $\sim$500 stars, assuming the  stars in the Solar Neighborhood are well mixed, though he worried about this assumption writing, p.259, that "The fact that for stars further than 100 parsecs to the north and to the south of the galactic plane there appears to be no trace of systematic relative motions lends some support to the assumption...that in the z-direction, the stars are thoroughly mixed".}. ESA’s Gaia mission \citep{gaia_prusti_2016,Lindegren_2018_gaiadr2} has changed this landscape: with 6D phase space and spectroscopy for millions of stars, we now observe Galactic dynamics from the inside. Newly revealed non-equilibrium structures are transforming and driving the theory of Galactic dynamics. A spectacular example of such structures is the unexpected phase spiral discovered in vertical phase space ($z, v_z$) by \citet{antoja_2018}, shown in Fig. \ref{fig:antoja}. This is the signature of on-going phase mixing \citep{tremaine_1999} in the vertical dimension of the disk after one or several perturbation(s), and indicates that our Galaxy is not in dynamical equilibrium. The phase spiral varies with position in the Galaxy, \citep{laporte_2019_sgrfootprint, bland-hawthorn_2019, li_2021, hunt_2022, frankel_2023, antoja_2023, alinder_2023, alinder_etal_2024}, age and metallicity  \citep{tian_etal_2018, laporte_2019_sgrfootprint, bland-hawthorn_2019, frankel_2025}, orbits \citep{li_2020}, and connects to global kinematics of the disk \citep{xu_etal_2020, bland-hawthorn_tepper-garcia_2021, gandhi_2022_snail}. 

Phase spirals can arise from multiple causes, such as interactions with external dwarf galaxies like the Sagittarius dwarf galaxy \citep[e.g.,][]{antoja_2018,binney_schoenrich_2018,laporte_2019_sgrfootprint, li_2020, bland-hawthorn_tepper-garcia_2021, li_2021,hunt_2021,banik_2022,banik_2023,asano_2025_ripples}, the impact of (dark) sub-halos \citep{chequers_2018,tremaine_2023,gilman_2025} or a DM subhalo such as the candidate recently identified from pulsar timing near the Sun \citep{Chakrabartietal2026}, wakes in dark matter (DM) halos \citep{grand_2022} or from internal perturbations such as spiral arms and the Galactic bar \citep{khoperskov_2019,hunt_2022,li_2023,chiba_2025_b}, a clumpy interstellar medium \citep{tepper-garcia_2025}, asymmetries in the halo of the Milky Way \citep[e.g. DM wakes][]{grand_2022}, or combinations of those \citep{garcia_conde2024}. Misaligned gas accretion can also create bending modes \citep[e.g.][]{kachaturyants_2022}. 

The phase spiral encodes information about when the perturbation started to phase-mix (the more tightly wound the phase spiral, the more time elapsed). Its amplitude must also be related to the strength of the perturbation that caused it. In addition, the shape of the spiral is determined by the Milky Way's gravitational potential, which is set by the mass distribution. Thus, the phase spiral constitutes an excellent tool not only to infer the recent history of the Milky Way, but also its global structure and mass distribution (e.g., \citealt{widmark_etal_2021b, widmark_2025, guo_etal_2022, guo_etal_2024}). 

However, no full understanding of this phenomenon can be achieved until missing physics in our models is incorporated. \citet{darling_widrow_2019}'s simulations showed that the phase spiral wound more slowly when self-gravity was included compared with massless particles. In addition, the relation between the dynamical perturbation in the vertical dimension and the other two (planar) dimensions in the disk is not fully understood. Interferences with other processes in the disk also remain under-explored. The discovery of two-armed spirals \citep{hunt_2022} in the inner disk of the Galaxy added complexity: one-armed spirals are related to bending waves, which are displacement of the mid-plane, while two armed spirals are believed to originate from breathing modes, which are expansions or contractions of vertical dimension of the disk \citep[e.g.,][]{widrow_2012, chiba_2025_b}, or the combination of two bending waves induced by multiple external perturbations \citep{lin_etal_2025}. Thus, our Galaxy may be exhibiting signals from multiple waves, with perhaps different origins, that we need to disentangle. 

This document aims to record and summarize the discussions that occurred during the meeting, identify future directions that will help understand the phase spiral in its details and in its context, set a reference point for future meetings, and exchange with the community more globally. It aims to be a snapshot of our thoughts at that point in time. For a recent and comprehensive review of the Galactic disk dynamics (and its phase spiral), we refer the reader to \cite{hunt_vasiliev_2025}.

\section{Aims Of The Workshop}
Seven years after that discovery, we wanted to bring the community together to discuss the current state of the field and its future directions. While the phase spiral has been followed up by about 50 articles of different teams, only by gathering together all the different skills and expertise (analytical modeling, simulations, data analysis, survey experts, dynamicists, etc.) related to the phase spiral, brought by a diverse group of participants, we would be able to advance fast in what has become one of the most challenging areas in the Milky Way research: the dynamics of disequilibria.

The questions that we wanted to address during the workshop are:
\begin{enumerate}\itemsep1pt
    \item What are the current theoretical and observational limitations in our understanding of phase spirals? 
    \item What is the origin(s) of galactic phase spirals?
    \item What physics, beyond phase mixing, is relevant to the evolution and morphology of phase spirals?
    \item  What are the connections between phase spirals and large-scale features of the disk, such as spiral structure, the bar, and the warp?
    \item What is the connection between halo substructure and phase spirals? Can we use phase spirals to learn something about the nature of dark matter?
\end{enumerate}
The workshop was organized along 5 themes that tackle these questions in an  intermixed way.

\section{Discussions \& Actions Items}

We briefly summarize the discussions that occurred at this workshop, focusing on recurrent topics and areas of progress. We refer the reader to the workshop webpage and discussion notes for a more accurate description of the discussions (see Section \ref{sec:resources}).

\subsection{The Origin Of The Phase Spiral}
The possible origins described in Section \ref{sec:intro} may produce similar signals. If those signals cannot be disentangled, the problem is partly or fully degenerate. Does the phase spiral alone have enough information to disentangle its possible origins? Suggestions for the way forward were to (1) articulate and simulate individual scenarios and combinations of them, (2) in parallel learn about the physical response of the disk to those perturbations, and (3) to maximize the extraction of the information content from the datasets on the Galactic disk by making a clear and detailed map\footnote{This is where J. Binney memorably suggests that `history should follow cartography', and we identified significant `cartography' issues, see Section \ref{sec:discrepancies}.}. Mapping the outer disk, which may be more sensitive to external perturbations, would also be helpful. In particular, it was shown that the Monoceros ring dominates in mass beyond 15 kpc and may be related to Sagittarius \citep[e.g.,][]{xu_newberg_carlin_2015}. Similarly, the Galactic warp exhibit higher modes that might be related to a recent excitation \citep{cabrera-gadea_2024}; all those structures may be connected to the phase spiral. 

\subsection{The Physics Of The Phase Spiral}

With an unknown origin, the main set of assumptions that could be made to describe the physics of the phase spiral is crucial. What are good initial conditions (equilibrium)? Is the physics involved too complex from the basic phase-mixing to `unwind' the phase spiral? Is the way forward to forward model the nonlinear processes that intervene in the dynamics of the disk? We were left with more questions than answers that we attempt to summarize below.

\topic{Self-gravity} is qualitatively well-understood, but has been quantitatively modeled in only a few experiments \citep{darling_widrow_2019, widrow_2023}. The stellar velocity dispersion in the disk, the presence of a bar and its resonances, such as the corotation at $\sim 7$ kpc, could affect the response of a self-gravitating disk to bending perturbations. It has been found to delay the onset of phase mixing, such that a phase spiral may look younger than it actually is \citep[by about 300 Myr in the solar neighborhood][]{asano_antoja_2025}.

\topic{The Interstellar medium} (ISM) has been shown to be relevant to understand the physics of the phase spiral in simulations \citep{tepper-garcia_2025}. The ISM can both dissolve the phase spiral via orbit-diffusion with local perturbations (e.g. with giant molecular clouds), or excite phase spirals via larger-scale perturbations \citep{tremaine_2023}. Modeling the full hydrodynamics at a resolution high enough is numerically challenging.

\topic{Warp, corrugations, flares} could be connected to the the phase spiral. Warps can arise from large-scale bending modes that wind into vertical corrugation patterns as a result of differential precession. Locally, this can manifest as a vertical phase spiral. The formation and evolution of these features have been modeled independently in the literature, with a variety of channels for their formation. We suggest that a more coordinated set of experiments modeling warp formation from a various sources, e.g., from misaligned gas accretion followed by an external perturbation (by, say, a satellite and/or DM halo), could help further understand the role(s) of externally-driven mechanisms for the phase spiral formation/evolution. 

\topic{Nonlinear effects} may be important given the multiple/periodic perturbations experienced by the disk. For example, when the disk is perturbed twice, it can exhibit three phase spirals, with the third arising from the nonlinear coupling between the first two, a phenomenon termed `galactic echo' \citep{chiba_2025_a}. Orbital resonance is another nonlinear effect that may drive phase spirals and may be particularly relevant for the formation of two-armed phase spirals \citep{li_2023,chiba_2025_b}.

\topic{Dark matter physics} could affect the presence, rate, and appearance of phase spirals. We discussed existing simulations with DM models, alternative to CDM, that could be explored.

\topic{The phase spiral is 6-dimensional.} We discussed avenues to model it in all phase space, in terms of coordinate systems and tools to extract the signal, such as? PCA, DMD, normalizing flows, action-angle variables (and zebra diagrams, $\theta_z-\Omega_z$), modeling the signal in $v_R, v_\varphi$ and abundances.

\topic{To progress on understanding the phase spiral}, we discussed a parallel framework where one branch would explore the theoretical aspects of the physics (e.g., self-gravity) and the other would forward model those aspects as accurately as possible to match the data, in order to empirically inform those forward models.

\subsection{Simulations} 

While simulations are not meant to reproduce the observed phase spiral in all its details, they enable us to test physical scenarios and validate measurements. We discussed their value in exploring the physical processes driving the disk's response to perturbations, and the origin(s) of the phase spiral.

\subsubsection{From Test Particle To Cosmological Simulations}
Simulations span from test particle simulations to full cosmological runs, each with different trade-offs, that we discussed in the context of the phase spiral.

\topic{Test particle simulations} evolve massless particles in a predetermined gravitational potential. They are computationally inexpensive, can match Gaia's resolution, and allow rapid exploration of parameter space \citep[e.g.,][]{li_2020, gandhi_2022_snail, li_2023, lin_etal_2025}, but lack self-consistency.

\topic{N-body and hydro simulations} with live stellar disks and, for the latter, gas \citep{hunt_2021, tepper-garcia_2025} account for both self-gravity and isolate effects like ISM-driven diffusion or bar resonances while retaining self-gravity. These are valuable for understanding how different physical processes individually and collectively shape the phase spiral \citep[e.g.,][]{laporte_2019_sgrfootprint}, but are costly to run and do not match Gaia's resolution. We discussed various acceleration techniques, such as GPU-accelerated N-body and hydro simulations \citep[e.g.,][]{hunt_2021,wibking_2022_quokka, david-cleris_2025_shamrock}, as well as machine-learning accelerated hydrodynamics \citep{hirashima_2025}.

\topic{Zoom-in cosmological simulations} like Auriga \citep{grand_etal_2017} re-simulate individual Milky Way-mass halos at high resolution within a cosmological context. These simulations capture emergent phenomena such as bars, warps, and satellite interactions that arise naturally from galaxy assembly. Although zoom-in cosmological simulations are far more computationally expensive than N-body and test-particle models at comparable resolution, some runs now reach the resolution required to produce phase spirals \citep{garcia_conde_2022,grand_2023}. 

\topic{The main questions} that were raised overlapped with other discussion groups too. In particular,
\begin{itemize}\itemsep-1mm
    \item What numerical resolution is needed for the simulation to reflect the physics governing the phase spiral and its evolution?
    \item What role do chaos and diffusion (physical and numerical) play? \cite{tremaine_2023} has shown that diffusion can affect the phase spiral significantly as it can erase structures at certain scales while preserving other scales for a longer time.
    \item What can be learned from simulations that do not account for self-gravity, such as test particle simulations?
    \item Would it be useful to create a set of simulations that vary the scenarios but keep the physical model fixed?
\end{itemize}

\subsubsection{Deliverable: A Public Simulation Database \label{subsubsection:simulation_database}}
We produced a curated simulation database with relevant contextual information (name, type, reference, link if public, resolution, and phase spiral-related information). A subset is shown in Table \ref{tab:sim_database} and the full table is a living document available at \href{https://docs.google.com/spreadsheets/d/16NGkLxDz0vY489BT1S7P_V63Wcua73WXC9tHl9RCktM}{this link}. We intend this to become a resource for the community, and encourage new simulations to be added through this {\protect{\href{https://docs.google.com/forms/d/e/1FAIpQLSdke5i-pL3YsNbxoMrkOaRrZC0vTMX7BC5cwNRuFaCQTcEeAQ/viewform}{web form}}).

\begin{table}[]
    \centering
    \begin{tabular}{|c|c|c|}
    \hline
      Reference & Resolution  & Public?\\
      \hline
       \cite{hunt_2021}  & 180 $M_\odot$ & Y\\
      \cite{asano_2025_ripples} & 180 $M_\odot$ & Y \\
       \cite{grand_etal_2017} & 50000-6000 $M_\odot$ & Y\\
       \cite{laporte_2018_sgr} & $10^7$ particles & N \\
       \hline
    \end{tabular}
    \caption{Subset of the simulation database. A living document is accessible at {\protect\href{https://docs.google.com/spreadsheets/d/16NGkLxDz0vY489BT1S7P_V63Wcua73WXC9tHl9RCktM}{this link}}.}
    \label{tab:sim_database}
\end{table}

\subsection{Quantifying The Phase Spiral}

\subsubsection{Measurements Discrepancies \label{sec:discrepancies}} 
The talk on \textit{Tools and Simulations} reviewed the different measurement methodologies and results attempting to \textit{quantify} the phase spiral. We shared and compiled literature measurements of the phase spiral properties. We compared the measurements of the amplitude of the phase spiral (its density contrast, $\delta f/f$ where $f$ is the phase space density) and the dynamical time associated with it (the time it would take the spiral to unwind in a static MW-like potential), as a function of angular momentum $L_z$. Specifically, we gathered results from \cite{antoja_2023, darragh-ford_2023, alinder_2023, frankel_2023, widmark_2025, frankel_2025} and Guo et al. (in prep). We found that while the results have qualitative resemblance with a wave-like pattern of those quantities with $L_z$ and similar variations, their values differ, and each study's patterns occur at different $L_z$. This result have led us to discuss the main causes for these discrepancies. Those can be differences in (1) the data used, (2) the parameter's definitions, (3) modeling of either the snail, the selection for the dataset, or both. We discussed producing a common data sample for phase spiral measurements and assessing desired properties. \textbf{What phase spiral properties do we want to measure?} Is the `dynamical time' well-defined and physically meaningful in the presence of non-linear effects?

\subsubsection{A Systematic Way to Characterize The Spiral} 

Science should yield reproducible results. We discussed ways to detect and characterize the phase spiral robustly. On the one hand, spiral detections in the literature have been done visually \citep{antoja_2018}, via an edge-detection technique \citep{antoja_2023} or via Fourier analysis \citep{garcia_conde_2022, frankel_2023}. On the other, the existence of the pattern resembling a two-armed spiral, or two one-arm spirals, in the inner disk of the Milky Way, justifies the needs for a more consistent methodology. 

Phase spiral measurements have mostly been parametric, oftentimes involving an amplitude, a dynamical time, and a phase. Within the literature, there are important differences between the definitions of those three parameters that might even not be physically meaningful. For example, the dynamical time is often used as `the time it would take to unwind the phase spiral in a static potential', whose meaning blurs in the presence of self-gravity and non-linear effects \citep{darling_widrow_2019,widrow_2023, asano_2025_ripples}, and it relies on an assumed vertical frequency $\Omega_z(J_\varphi, J_z)$ and thus on the gravitational potential $\Phi_\mathrm{pot}$.

These arguments have led us to consider a more mathematical (and less physical) representation of the pattern that could be equally used for simulations and for observations. Models that have been discussed are the Archimedean spiral, as well as parameter-free spiral detection and characterization using modern and tailored machine learning techniques. Other promising examples are basis function expansions, which provide quantitative representations of the phase spirals and can easily separate one-armed and two-armed features from each other and from the equilibrium background (Tavangar et al., in prep.). More broadly, we discussed the benefits and tradeoffs of models with various levels of interpretability, physical realism, and those requiring human/machine decision making, without reaching a clear consensus.

\subsubsection{Deliverable: A Public Standardized Sample \label{subsubsection:golden_sample}}

Any modeling, inference, or conclusion drawn from a dataset relies on a set of subjective choices made during dataset and model construction\footnote{D.W.Hogg KoCo, MPIA, 2025-08-01, \it{When is a measurement not really a measurement?}}. Measurements using different techniques and data samples yield different results, but it is unclear to what extent the discrepancies arise from the methodology or from the dataset.

We suggested to produce a public standardized sample. The goal would be two-fold; (1) ensuring to the best we can a sample with the quality requirements to allow the analysis of the phase spiral, limiting biases and uncertainties, and with a well-defined selection function (that would also be provided, see below); and (2) providing a sample that enables different methodologies to compare to each other quantitatively. A forward model of the selection function that can be evaluated in a simple way is necessary for proper data-model comparisons. It would also be released with those curated samples, most likely involving the GaiaUnlimited framework \citep{cantat-gaudin_2023}. This would also require a strategy express the selection function's input variables ($l, b, m, C$) as a function of the variables used by models, such as Galactocentric positions and absolute magnitudes.

While the ideal sample covers a large surface area of the disk in 6D to address the questions about the physics and the origin of the phase spiral (to produce a map) and be representative of the full underlying population of stars (to trace gradients or chronology), we concluded that each specific scientific question would probably require its own specific dataset. However, for the sake of reproducible science, we argued for measurements to be made anyway on one or two of those benchmark datasets, in order to help the community understand the differences between their findings.

We propose to construct two samples: a 6D local sample that should be as simple as possible, and a 5D one (without radial velocities) with a larger spatial coverage of the disk (until future datasets can be both large and 6D). This paragraph only reflects our current position on the construction of such a standardized sample, but it may evolve over the course of future experimentation.

\topic{For the local 6D sample}, we propose the following qualitative criteria in the data
\begin{itemize}\itemsep-1mm
    \item high quality cuts in parallaxes, radial velocities and proper motion, such as $\sigma_\varpi / \varpi < 10\%$, $\mathrm{RUWE < 1.4}$
    \item a distance cut to target the solar neighborhood (e.g. $D\leq 500$ pc)
    \item while conserving a large number of sources to fully represent the local 6D phase space.
\end{itemize}

\topic{For the global 5D sample}, we discussed the following qualitative criteria
\begin{itemize}\itemsep-1mm
    \item should cover a large surface area of the disk
    \item would leverage the large distances of stars in the disk, for which the vertical motion $v_z$ can be mostly determined from the proper motion
    \item might need to be strategic so that its overlap with resonances with the bar.
\end{itemize}
To be more quantitative, a working group started to use simulations to predict the effect of data uncertainties on phase spiral detection and measurements. We also discussed which data columns to use: should the catalog offer distance estimates and/or parallaxes? Which distance estimates? Should it produce Galactocentric coordinates or let the community choose which Galactic parameters to assume?

Finally, we discussed the roles of surveys other than Gaia and how they (will) contribute to understanding the phase spiral. Those include Skymapper \citep{skymapper_ref_2007}, GaiaNIR \citep{hobbs_2016}, and SDSS-V \citep{Kollmeier_2026_SDSSV}, although we lacked the relevant expertise locally to make it an immediate actionable item. The community working on the phase spiral would, however, benefit from input from and collaborations with the broader community.

\section{Summary \& Future Directions}
This meeting addressed research questions about the Gaia Phase Spiral from various domains (theory, simulations, observations) and at different scales (from details of parameters to the bigger picture). While those discussions were not expected to, and did not reach, a global consensus, several directions have been identified. This document aims to be a snapshot in time of on-going knowledge and lack there-of on this topic.

\topic{Controlled simulations} with a fixed physical model and different perturbation origins can provide helpful constraints on the origin of the phase spiral, making comparisons between scenarios more apple-to-apple than the current literature. Controlled simulations with a fixed perturbation origin and different physical models can instead shed light on the physics governing the phase spiral, with simple experiments varying resolution, gas physics, and DM models helping to isolate and rank different effects. Cosmological simulations with fixed or carefully varied initial conditions can provide added realism to idealized experiments. A set of hybrid ``cosmo-idealized'' runs could explore regions of model physics space identified from controlled simulations, potentially extracting trends with, e.g., stellar age.

\topic{Accurate description and inference of the phase spiral} in the observations and in simulations can help extractive quantitative information from the phase spiral. When compared to simulations thoroughly, accounting for selection effects and observational uncertainties, such statistical models could be powerful in helping constrain the theory against our real Milky Way, rule out some scenarios, and extract the physics driving the response of the disk to perturbations.

\topic{Maximizing the use of the information content from datasets}, e.g. by mapping the 6-dimensional perturbation (not just the local $z-v_z$ plane would be extremely useful in disentangling between various theories. This would require simple perturbation models to go 6D, and a more complex treatment of datasets. For example, it could require joining several different datasets from various surveys, or using all of Gaia (where some regions of the parameter space have no radial velocity or large astrometric uncertainties). 

\topic{To start in the directions outlined above}, we have begun compiling (1) a curated list of simulation outputs for studying disk perturbations (Section \ref{subsubsection:simulation_database}), (2) a Gaia golden sample for testing and comparing inference techniques (Section \ref{subsubsection:golden_sample}), (3) a comparison of current phase spiral measurement discrepancies and their origins (Section \ref{sec:discrepancies}), and (4) a mentoring network to foster collaborations and support junior scientists in this subfield (Section \ref{sec:resources}).

\section{Resources\label{sec:resources}}
\begin{itemize} \itemsep-0.5mm
    \item \href{https://sites.google.com/fqa.ub.edu/phasespiralworkshop/home?authuser=0}{Webpage of the workshop}
    \item \href{https://docs.google.com/document/d/1dq73b6FYDFK6oBChxcWRbZoeh4uSeeq-M88R84c9XTs/edit?usp=sharing}{Notes taken during the workshop} 
    \item \href{https://drive.google.com/drive/folders/1dQevvGTuljWI8MPRyt5szMiboIOEz9pJ}{Slides used during the workshop}
  \item \href{https://docs.google.com/spreadsheets/d/16NGkLxDz0vY489BT1S7P_V63Wcua73WXC9tHl9RCktM}{Link to simulation database}
    \item 
    Mentoring network: Researchers of all levels who work on the phase spiral are invited to join as a mentor, mentee, or both! Existing mentoring networks in other fields have frequently yielded positive results and new collaborations, and we hope the same will be true for the phase spiral community. One can join by filling out this \href{https://forms.gle/N7Af1p5vcVEHfraT8}{google form}.
\end{itemize}

\section*{Acknowledgments}
This document was initiated and partly written at the Winding, Unwinding, Rewinding the Gaia Phase Spiral workshop, hosted by the Lorentz Center. All participants to this workshop contributed valuable inputs into the workshop, the discussions, the presentations, and the meeting notes that this summary is based on. The authors of this document hereby thank all participants of the workshop.
NF acknowledges the support of the Natural Sciences and Engineering Research Council of Canada (NSERC), funding reference numbers 568580 and RGPIN-2020-03885, and partial support from an Arts \& Sciences Postdoctoral Fellowship at the University of Toronto. 
MS acknowledges support from the Spanish MICIN/AEI/10.13039/501100011033 and "ERDF A way of making Europe" by the “European Union” and the European Union "Next Generation EU"/PRTR, through grants PID2021-125451NA-I00 and CNS2022-135232, and the Institute of Cosmos Sciences University of Barcelona (ICCUB, Unidad de Excelencia ’Mar\'{\i}a de Maeztu’) through grant CEX2019-000918-M. 
SK acknowledges support by the Deutsche Forschungsgemeinschaft under the grant KH~500/2-1.
ZYL is supported by the National Natural Science Foundation of China under grant Nos. 12233001, 12533004, by  the National Key R\&D Program of China under grant No. 2024YFA1611602, and by a Shanghai Natural Science Research Grant (24ZR1491200).

\bibliographystyle{aasjournalv7}
\bibliography{lit}

\end{document}